\documentclass[runningheads]{svmult}
\usepackage{graphicx}  
\usepackage{epsf} 

      \def\bpkm{B^-\rightarrow \phi K^-} 
\def\beq{\begin{equation}} \def\eeq{\end{equation}}
\def\beqa{\begin{eqnarray}} \def\eeqa{\end{eqnarray}}
\def\ba{\begin{array}} \def\ea{\end{array}}

\input epsf.tex
\def\DESepsf(#1 width #2){\epsfxsize=#2 \epsfbox{#1}}

\begin{document}
\title*{Dark Matter, Muon g-2 and Other SUSY Constraints}
\toctitle{Dark Matter, Muon g-2 and Other SUSY Constraints}
%
%
\titlerunning{Dark Matter, Muon g-2 and Other SUSY Constraints}
%
\author{Richard Arnowitt\inst{1}
\and Bhaskar Dutta\inst{2}
\and Bo Hu\inst{1}}
\authorrunning{Richard Arnowitt et al.}
%
%
\institute{Center For Theoretical Physics, Department of Physics,\\
					 Texas A\&M University, College Station, TX 77843-4242, USA
\and Department of Physics, University of Regina, Regina SK, S4S 0A2 Canada}

\maketitle

\begin{abstract}
Recent developments constraining the SUSY parameter space are reviewed
within the framework of SUGRA GUT models. The WMAP data is seen to reduce
the error in the density of cold dark matter by about a factor of four,
implying that the lightest stau is only 5 -10 GeV heavier than the lightest
neutralino when $m_0$, $m_{1/2} < 1$ TeV. The CMD-2 re-analysis of their data has
reduced the disagreement between the Standard Model prediction and the
Brookhaven measurement of the muon magnetic moment to 1.9 $\sigma$, while
using the tau decay data plus CVC, the disagreement is 0.7 $\sigma$. (However,
the two sets of data remain inconsistent at the 2.9 $\sigma$ level.) The
recent Belle and BABAR measurements of the $B\rightarrow \phi K$ CP violating
parameters and branching ratios are discussed. They are analyzed
theoretically within the BBNS improved factorization method. The CP
parameters are in disagreement with the Standard Model at the 2.7 $\sigma$
level, and the branching ratios are low by a factor of two or more over
most of the parameter space. It is shown that both anomalies can naturally
be accounted for by adding a non-universal cubic soft breaking term at $M_G$
mixing the second and third generations.
\end{abstract}

\section{Introduction}

While SUSY particles are yet to be discovered, a wide range of data has begun to
limit the allowed SUSY parameter space. We review here what has happenned over
the past year to further restrict SUSY models of particle physics. A number of
new experimental and theoretical analyses have occurred:

The current experiments that most strongly restrict the SUSY parameter
space are the following:
\begin{itemize}
\item WMAP data has greatly constrained the basic cosmological parameters
\item While the analysis of $\mu^-$ data for the muon anomalous magnetic moment
has not yet been completed, there has been further experimental results and
theoretical analysis that have modified the theoretical Standard Model (SM)
prediction of $g_{\mu}-2$.
\item The B-factories, BABAR and Belle measurements of the CP violating B decays,
particularly $B^0\rightarrow J/\psi K_s$ and $B^0\rightarrow\phi K_s$,
$B^\pm\rightarrow\phi K^\pm$ impose new constraints on any new theory of CP
violation when they are combined with theoretical advances that have occurred in
calculating these decays.
\end{itemize}

In addition one must continue to impose the previously known constraints on the
SUSY parameter space. The most important of these are:
\begin{itemize}
\item The light Higgs mass bound $m_h>$114.1 GeV \cite{higgs1}
\item The $b\rightarrow s\gamma$ branching ratio constraint \cite{bsg1}
\item The light chargino mass bound $m_{\tilde\chi^{\pm}_1}>103$ GeV \cite{lep}
\item The electron and neutron electric dipole moments bounds \cite{PDG}
\end{itemize}

In order to analyze these phenomea it is necessary to chose the SUSY model. At
one extreme one has the MSSM with over 100 free parameters (with 43 CP violating
phases). At the other one has mSUGRA with four parameters and one sign (often
augmented with two or four additional CP violating phases). The large number of free
parameters in the MSSM generally allows one to fit each experimental constraint
seperately by tuning one or another parameter. For mSUGRA, where the parameters
are specified at the GUT scale, $M_G\cong 2\times 10^{16}$ GeV, the situation is
far more constrained. One can ask here if all the data taken together is still
consistent with mSUGRA (or indeed with the Standard Model). The new B-factory
data appears to be putting the greatest strain on mSUGRA, and if the current data
is confirmed, it may be that one is seeing a breakdown of mSUGRA (and also the
SM) for the first time and one may need a modification of the universal soft
breaking assumptions of mSUGRA.  

\section{mSUGRA Model}
\begin{figure}[ht]\vspace{-0cm}
\centerline{ \DESepsf(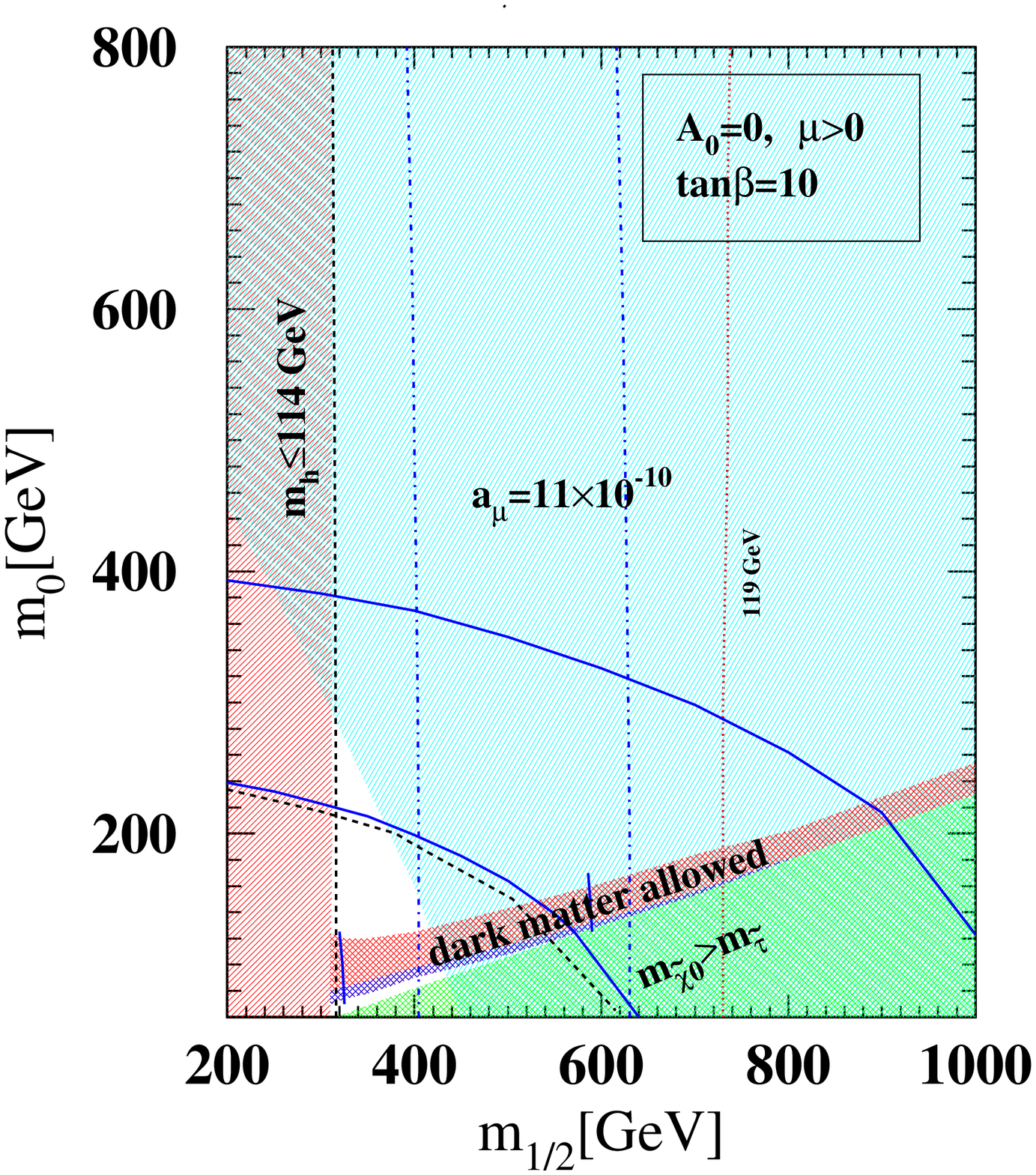 width 8 cm) }
\caption {\label{fig1}  Allowed region in the $m_0$ - $m_{1/2}$ plane from the relic density 
constraint  for $\tan\beta = 10$, $A_0  = 0$ and $\mu >0$ \cite{t10}. The red region was 
allowed by the older balloon data, and the narrow blue band by the new WMAP 
data. The dotted red vertical lines are different Higgs masses, and the 
current LEP bound produces the lower bound on $m_{1/2}$. The light blue region 
is excluded if $\delta a_{\mu} > 11 \times 10^{-10}$. (Other lines are discussed in
reference \cite{t10}.)} 
\end{figure}

\begin{figure}[t]
\vspace{-0cm}
\centerline{ \DESepsf(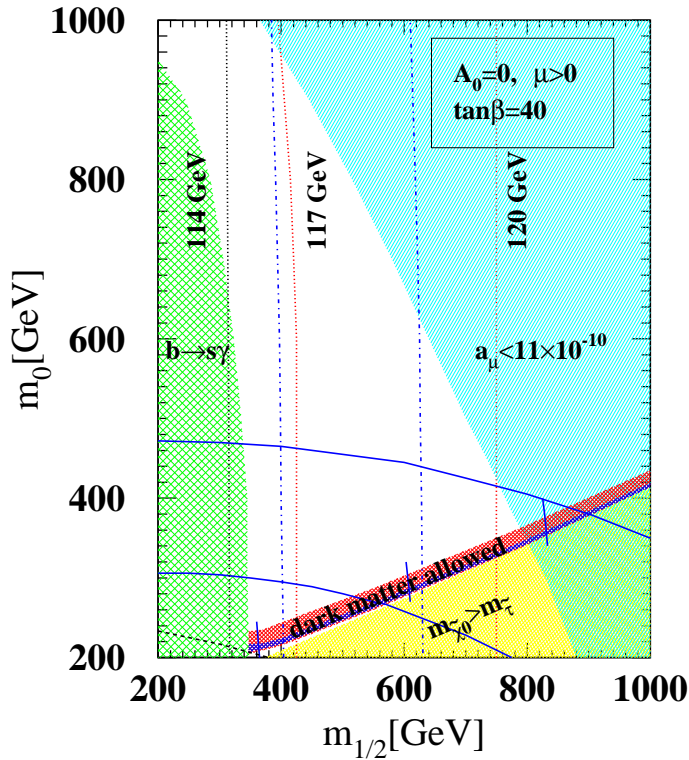 width 8 cm) }
\caption {\label{fig2}  Same as Fig. 1 for $\tan\beta = 40$, $A_0 =0$, $\mu >0$ except that now that 
the $b \rightarrow s  \gamma$ constraint (green region) produces the lower bound on
$m_{1/2}$ \cite{t10}.} 
\end{figure}

 \begin{figure}[t]\vspace{-0cm}
\centerline{ \DESepsf(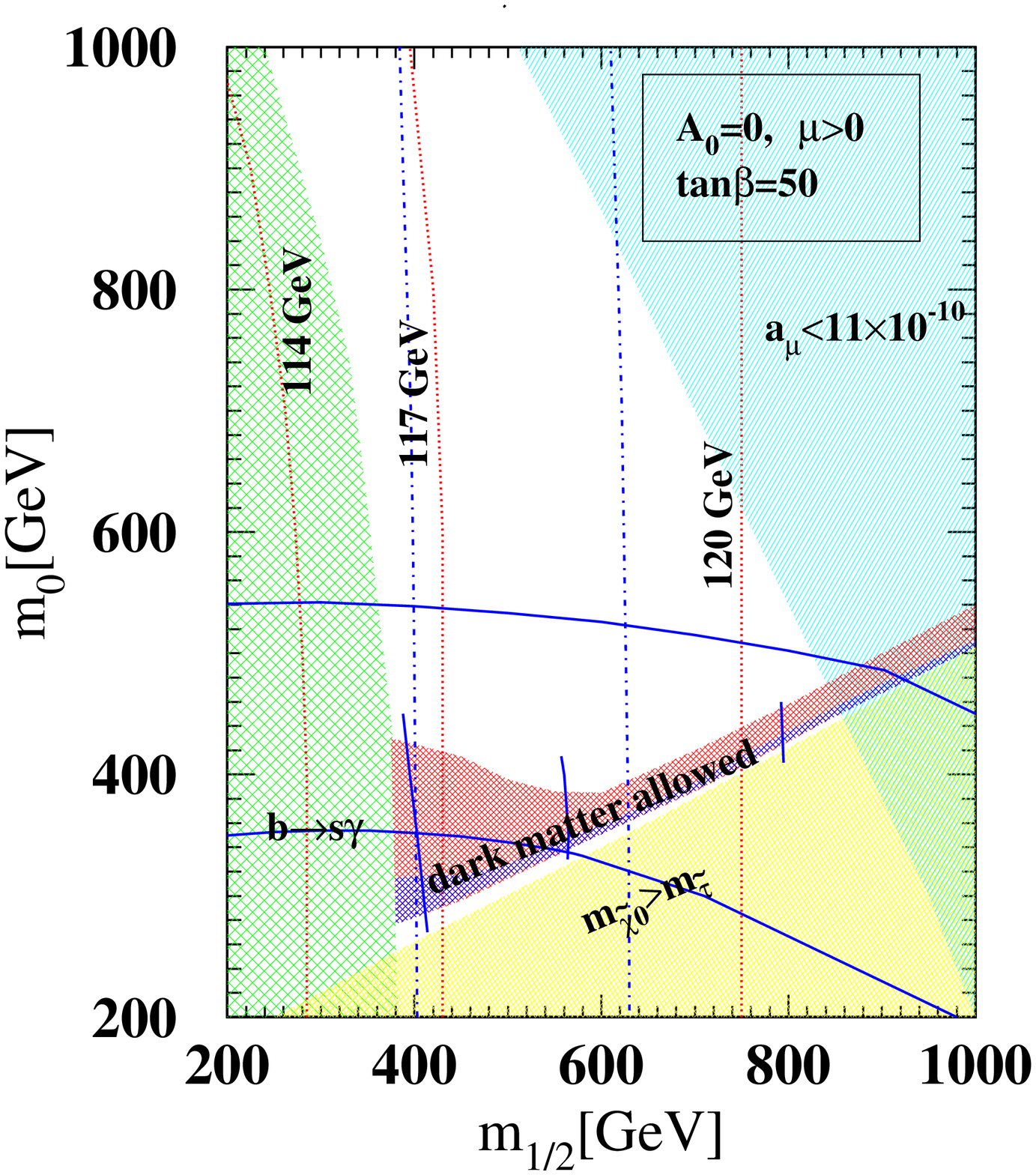 width 8 cm) }
\caption {\label{fig3}  Same as Fig. 2 for $\tan\beta = 50$, $A_0$ = 0, $\mu > 0$. Note that the 
large bulge at lower $m_{1/2}$ allowed by the older balloon data is now mostly 
excluded by the WMAP data \cite{t10}.} 
\end{figure}
We begin by reviewing the status of mSUGRA models with R-parity invariance as of
a year ago. Recall that mSUGRA depends on the following parameters: $m_0$, the
universal soft breaking mass at $M_G$; $m_{1/2}$, the universal gaugino mass at
$M_G$; $\tan\beta=\langle H_2 \rangle/\langle H_1 \rangle$ where $\langle H_{1,2} \rangle$ gives rise to $(d,\, u)$ quark
masses. In addition, the model allows $A_0$ and the $\mu$ parameter to be complex
(where $\mu$ is the Higgs mixing parameter in the superpotential term $\mu H_1
H_2$). The renormalization group equations (RGE) leading to electroweak
symmetry breaking at the electroweak scale determine $|\mu|$ and so there can be
two CP violating phases, $\mu=|\mu|e^{i\alpha_\mu}$ and $A_0=|A_0|e^{i\phi}$ (in addition to the CKM CP
violating phase). To accomodate the electron and neutron electric dipole moments we will
also allow the three gaugino masses at $M_G$ to have phases: $\tilde m_i=m_{1/2}
e^{i\phi_i}$ with the convention $\phi_2=0$.

The allowed parameter space for $m_0,\, m_{1/2}<1$ TeV is shown in Figs.~1, 2, 3
for $\tan\beta=10,\, 40$, 50 with $A_0=0$, $\mu>0$ ($\mu$ real)\cite{t10}. Here the
lightest neutralino $\tilde\chi^0_1$ is the dark matter candidate, and the narrow
rising (pink) band is the region of parameter space where the predicted amount of
relic dark matter left over after the Big Bang is in agreement with the CMB
measurements of
$\Omega_{DM}=\rho_{DM}/\rho_c$ as of a year ago from the various balloon flights. 
Here $\rho_{DM}$ is the dark matter (DM) mass
density, and $\rho_c$ is the critical density to close the universe ( $\rho_c=3
H_0^2/8\pi G_N$, $H_0$=Hubble constant, $G_N$=Newton constant)

It is important to realize that the narrowness of the dark matter allowed band is
{\it not} a fine tuning but rather a consequence of the co-annihilation effect
for $\tilde\chi^0_1$ and the light $\tilde\tau_1$ in the early universe. This
arises naturally in mSUGRA (and are generic features for many GUT models) due to
the near accidental degeneracies between the $\tilde\tau_1$ and the
$\tilde\chi^0_1$ and the fact that the Boltzman exponential factors in the early
universe annihilation analysis produces sharp cut offs in the relic density. Thus
the bottom of the allowed band is at  the experimental lower bound on
$\Omega_{DM}$ (where $m_{\tilde\tau_1}$-$m_{\tilde\chi^0_1}$ takes its minimum
allowed value) and the top corresponds to the experimental upper bound on
$\Omega_{DM}$ (above which $m_{\tilde\tau_1}$-$m_{\tilde\chi^0_1}$ is too large
to get efficient early universe annihilation).

If one allows $m_0$ and $m_{1/2}$ to be in the multi-TeV region, two additional
region occur with acceptable relic density \cite{nath}: (1) the ``focus point"
region where $m_0\stackrel{>}{\sim} 1$ TeV and $m_{1/2}\stackrel{>}{\sim}$ 400 GeV and (2) the funnel region where
$m_0\simeq m_{1/2} \stackrel{>}{\sim}1$ TeV and $\tan\beta$ is large. These have been studied by a
number of authors (e.g. \cite{nath,foc1,foc2,foc3,foc4,foc5,foc6,foc7,foc8}). If the $g_{\mu}-2$ data eventually confirms a
deviation from the SM,  would be eliminated (as shown in the blue regions of
Figs. 1-3) but this is still in doubt as we discuss below. However, aside from
$g_{\mu}-2$, these are regions of relatively high fine tuning. Thus one can define
the fine tuning parameter 
\begin{equation}
\Delta_\Omega\equiv
\left[\sum_i{{\partial ln\Omega_{\tilde\chi^0_1}h^2}\over{\partial ln
a_i}}\right]^{1/2};\,a_i=m_0,\,m_{1/2},\, etc.
\end{equation} and $H_0=h$(100 km/sec Mpc).
Large $\Delta_\Omega$ implies significant fine tuning.
Figs 4 and 5 \cite{Ellis} show the values for $\Delta_\Omega$ for the focus-point region
($\tan\beta=10$) and the funnel region $\tan\beta=50$. One sees that in the
co-annihilation region  $\Delta_\Omega\simeq 1-10$, while in the focus point  or
funnel regions $\Delta_\Omega\simeq 100-1000$. Whether this
is an argument to exclude these high fine tuning regions is a matter of taste.
\begin{figure}[t]\vspace{-0cm}
\centerline{ \DESepsf(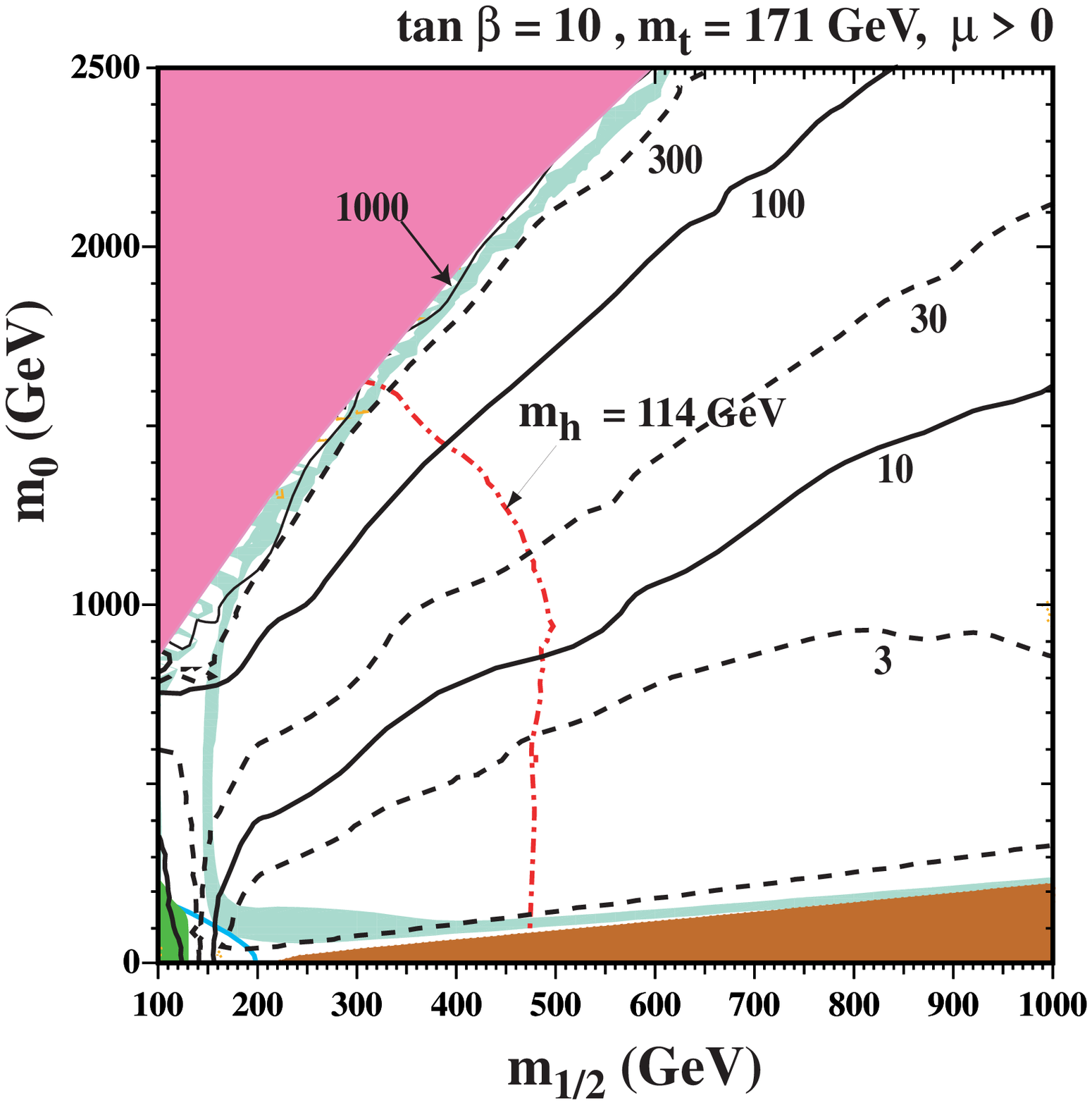 width 8 cm) }
\caption {Contours of total sensitivity $\Delta^{\Omega}$, $\tan\beta=10$, $A_0=0$, $\mu>0$,
$m_t=171$ GeV \cite{Ellis}.\label{fig5} } 
\end{figure}
 \begin{figure}[t]\vspace{-0cm}
\centerline{ \DESepsf(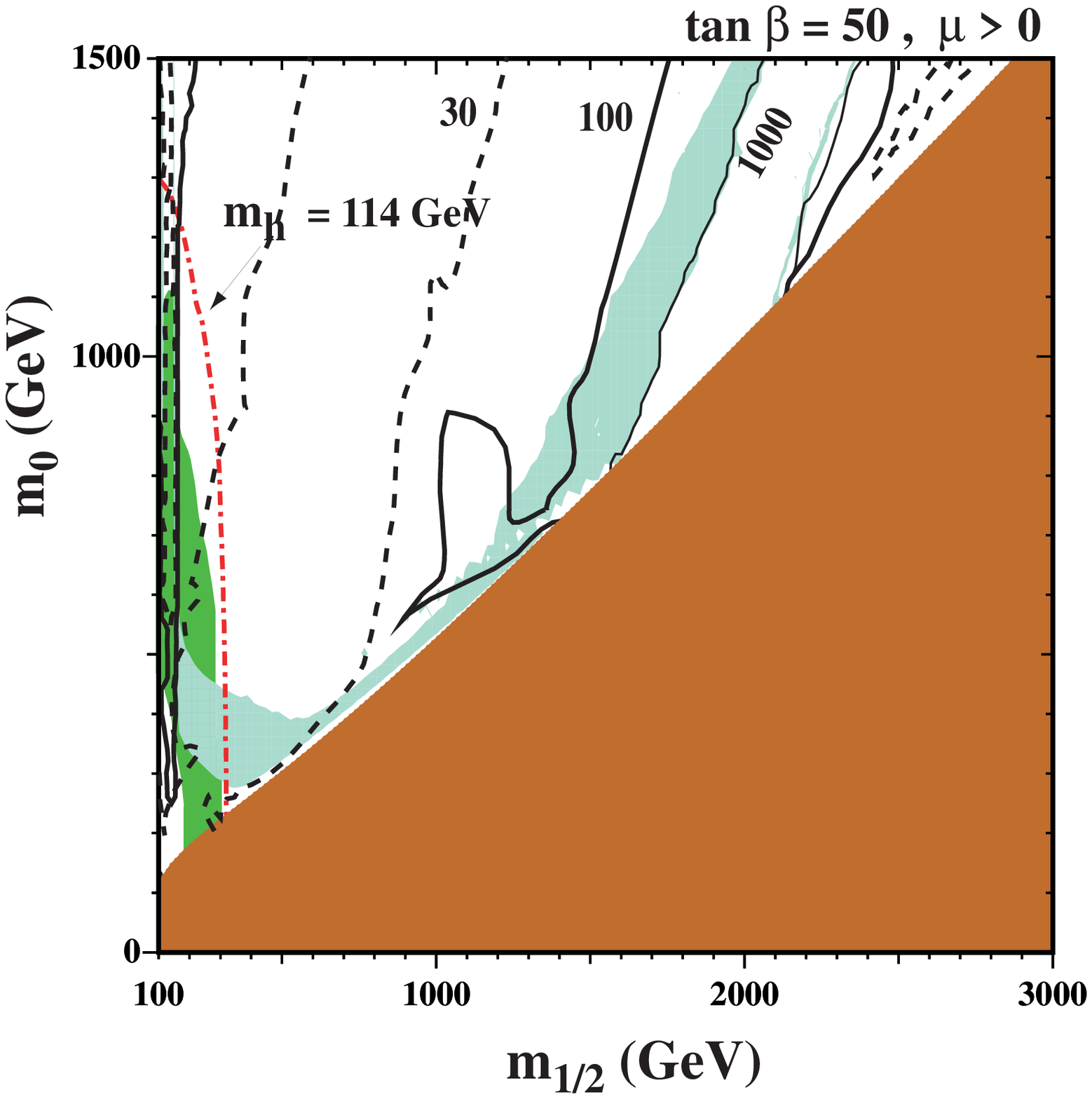 width 8 cm) }
\caption {\label{fig4}Contours of total sensitivity $\Delta^{\Omega}$, $\tan\beta=50$, $A_0=0$, $\mu>0$,
$m_t=175$ GeV \cite{Ellis}.} 
\end{figure}

\section{The WMAP data}
The Wilkinson Microwave Anisotropy Probe (WMAP) has determined the basic
cosmological parameters with remarkable precesion. What is measured is the baryon
density $\Omega_m$, the total density $\Omega_{\rm tot}$, the Hubble constant
$h=H/100(km/sec Mpc)$, as well as many other cosmological parameters.
Characterizing the difference between $\Omega_{\rm tot}$ and $\Omega_m$ by a
cosmological constant $\Lambda$, WMAP finds $\Omega_b=$0.044$\pm$0.004,
$\Omega_m=$0.27$\pm$0.04, $\Omega_\Lambda=$0.73$\pm$0.04,
$h=0.71^{+0.04}_{-0.03}$. $\Omega_\Lambda$ is in good agreement with the direct
measurement of $\Lambda$ using type IA supernovae \cite{sup1} and $h$
agrees very well with the Hubble Key Project's direct measurement \cite{h1}.
$\Omega_{\rm CDM}$ is then given by $\Omega_m-\Omega_b$, and there is a general
concordance with a cold dark matter universe with a cosmological constant. (The distinction
between a cosmological constant and  and a
quintessence model is not yet measurable). The WMAP $2\sigma$ range for
$\Omega_{\rm CDM}$ is 
\begin{equation}
0.094<\Omega_{\rm CDM}h^2<0.13
\end{equation}
This is to be compared with the previous balloon flight measurements of 
$0.07<\Omega_{\rm CDM}h^2<0.21$. Thus WMAP has reduced the uncertainty in the
amount of cold dark matter by nearly a factor of four! This can be seen in
Figs.1-3, where the WMAP allowed region is the narrow blue band. Particularly
striking is the reduction of parameter space for $\tan\beta\stackrel{>}{\sim}45$ as seen
in Fig.3. In addition an upper bound $m_{1/2}\stackrel{<}{\sim}1$ TeV is found in
\cite{foc6}. As the data becomes more and more accurates the CMB measurements
will effectively determine one of the mSUGRA parameters i.e. determine a
relation of the form (neglecting CP violating phases)\begin{equation}
m_0=m_0(m_{1/2},\, A_0,\,\tan\beta,\,\mu/|\mu|)
\end{equation}
\section{Update on $g_\mu$-2}
While there has been no new data this year on the muon magnetic moment anomaly
$a_\mu=1/2(g_\mu-2)$, there has been a re-evaluation of some of the old data used
to calculate the SM contribution, which has reduced the significance of the
effect. We begin first by reviewing  where things stood last year. First, the
current experimental values of the electron \cite{emu} and muon \cite{mmu}
anomalies are
\begin{equation}
a_{e}^{\rm exp}=1159652188.3(4.2)\times 10^{-12}
\end{equation}
\begin{equation}
a_{\mu}^{\rm exp}=11659203(8)\times 10^{-10}
\end{equation}
The theoretical values of $a_e$ can be expressed as a power series in
$\alpha$ \cite{ae} 
\begin{equation}
a_{e}^{\rm th}=\sum_i^4 S^e_n({\alpha\over \pi})+1.70(3)\times 10^{-12}
\end{equation}
where $S^e_n$ begins as $S^e_1=1/2$ (the Schwinger term) and the last term in
Eq.(6) are the electroweak and hadronic contributions. One equates Eqs.(4) and
(6) and solves for $\alpha$. This is currently the most accurate detremination
of the fine structure constant at zero energy: 
\begin{equation}
\alpha^{-1}=137.03599875(52)
\end{equation}
This number is about 1.6$\sigma$ down from the previously given values due to
the correction of an error recently found in the $\alpha^4$ term \cite{ae0}.

The SM prediction for $a_{\mu}$ can be divided into QED, electroweak and
hadronic parts:
\begin{equation}
               a_\mu^{\rm SM} = a_\mu^{\rm QED} + a_\mu^{\rm EW} + a_\mu^{\rm had}
\end{equation}
The QED part has been calculated through order $\alpha^5$. Fortunately, the new
value of $\alpha$ does not change this significantly \cite{ae0}:
\begin{equation}
a_{\mu}^{\rm QED}=11658470.35(28)\times 10^{-10}
\end{equation}
The electroweak contribution has been  calculated to two
loop order (the two loop part being surprisingly large) by two
groups\cite{ae1,ae2} with good agreement yielding an average
\begin{equation}
a_{\mu}^{\rm EW}=15.3(0.2)\times 10^{-10}
\end{equation}
The hadronic contribution can be divided into three parts, a leading order, a
higher order and a scattering of light by light (LbL) part. The latter two have
been done by several groups, and there is now reasoable agreement in the LbL
part:\begin{equation}
a_{\mu}^{\rm had HO}=-10.0(0.6)\times10^{-10};\,a_\mu^{\rm LbL} =
8 (4)\times10^{-10}\end{equation}

The term which is most unclear is the leading order hadronic contribution. This
can be calculated by using a dispertion relation:
\begin{equation}
  a_\mu^{\rm LO} ={\alpha^2\over {3\pi^2}}\int^{\infty}_{4m^2_{\pi}}ds{K(s)\over s}
  \left[{{\sigma^{(0)}(e^+e^-\rightarrow{\rm hadrons})}\over{\sigma(e^+e^-\rightarrow
\mu^+\mu^-)}}\right]
\end{equation}
In principle, one uses the experimental ${e^+e^-}\rightarrow $hadrons cross section to calculate the
integral which is dominated by the low energy part. Note that $\sigma^{(0)}_{e^+e^-}$ means the
experimental cross section corrected for initial radiation, electron vertex and
photon vacuum polarization (so one does not count the higher order
contributions).

A large number of experimental groups have contributed to the determination of
$\sigma^{(0)}_{e^+e^-}$ and two independent ways have been used:\\
\noindent  (i) Direct measurement of $\sigma({e^+e^-}\rightarrow hadrons)$ 
 with the above mentioned
radiative corrections made. This data has been dominated by the very accurate
CMD-2 experiment.
\\\noindent  (ii) One uses the 2$\pi$ decays of the $\tau$, which goes through the 
vector (V)
interaction, and then uses CVC to determine $\sigma_{e^+e^-}$ for 
$\sqrt s \stackrel{<}{\sim}1.7$ GeV
(the kinematic reach of the $\tau$ decay). Here one must make corrections to
account for the breaking of CVC. This data is dominated by ALEPH.

The results of these calculation, as of last year were very puzzling. Three
groups did the analysis using method (i) getting almost identitical
answers\cite{amu,amu1,amu2}. An average value was $a_{\mu}^{\rm had
LO}(e^+e^-)=(683.8\pm 7.0)\times 10^{-10}$. However, using the $\tau$ data
(augmented by the $e^+e^-$ data at higher energy), Ref.\cite{amu} found $a_{\mu}^{\rm had
LO}(e^+e^-)=(709.0\pm 5.9)\times 10^{-10}$. (In fact, the $e^+e^-$ and $\tau$
data were inconsistent with each other at the level of 4.6$\sigma$!) The $e^+e^-$ data
gave rise to a 3.0$\sigma$ disagreement between experiment and the SM while the
$\tau$ gave a smaller 1.0$\sigma$ disagreement.

Recently, however there have been new results that have significantly reduced
the discrepancy between the two approaches and between experiment and the SM.
CMD-2 found an error in its analysis of the radiative correction to be made to
their $e^+e^-\rightarrow$ had data (a lepton vacuum polarization diagram was
omitted). A full re-analysis of the SM value of $a_{\mu}^{\rm had LO}$ was then
carried out in \cite{amu4} with the results now of $a_{\mu}^{\rm had
LO}(e^+e^-)=(696.3\pm 7.2)\times 10^{-10}$ and $a_{\mu}^{\rm had
LO}(\tau based)=(711.0\pm 5.8)\times 10^{-10}$. The discrepancy with the SM now
becomes
\begin{equation}
a_{\mu}^{\rm exp}-a_{\mu}^{\rm SM}(e^+e^-)=(22.1\pm 11.3)\times 10^{-10}
\end{equation}
\begin{equation}
a_{\mu}^{\rm exp}-a_{\mu}^{\rm SM}(\tau)=(7.4\pm 10.5)\times 10^{-10}
\end{equation}
corresponding to 1.9$\sigma$ and 0.7 $\sigma$ descrepancy respectively. However
the $\tau$ data still disagrees with the $e^+e^-$ data at a 2.9$\sigma$ level
(the disagreement occuring particularly at $\sqrt s>850$ MeV) and so a puzzle
still remains. 

One may expect some further clarification in the not too distant future. Thus
the Brookhaven analysis of their $\mu^-$ data should reduce the experimental
error on $a_{\mu}^{\rm exp}$. Further, new data from KLOE and BABAR can check
the CMD-2 results. In Figs.(1-3), we have shown the excluded region in the
parameter space (blue) if the descripency with the SM is $11\times 10^{-10}$
(i.e.about 1$\sigma$ from zero). A large amount of parameter space is
eliminated.
\section{CP Violating B decays}
BABAR and Belle have now measured with increasing accuracy a number of CP
violating B decays. This has opened up new tests of the Standard Model and new
ways to search for new physics.

Simultaneously, improved techniques for calculating these decays have been
developed over the past two years by Beneke, Buchalla, Neubert and
Sachrajda \cite{bbns} (BBNS) and these procedures have been further discussed by Du et
al.~\cite{du}. In previous analyses using the so-called ``naive factorization",
decay amplitudes for $B\rightarrow M_1+M_2$ depend on the matrix elements of
operators $O_i$, $\langle M_1M_2|O_i| B \rangle$, and these matrix elements were factorized to
calculate them. In the BBNS scheme, ``non-factorizable" contributions can be
calculated allowing the calculations of the strong phase, which is needed to
discuss direct CP violation.

 We consider here the decays  $B^0\rightarrow J/\Psi K_s$, $B^0\rightarrow \phi
 K_s$ and $B^{\pm}\rightarrow \phi K^{\pm}$. The results below are presented using the BBNS
 analysis and we discuss here the decays for SUGRA models. We require, of
 course, the simultaneous satisfaction of the dark matter, the $b\rightarrow s
 \gamma$ branching ratio, the electron and neutron EDM constraints
 etc\footnote{These B decays have been previously examined in SUSY models by a number of
 different authors \cite{bdecay} in low energy MSSM using the mass insertion
 method \cite{mb} and/or without taking into full account of the BBNS analysis}.
 
 The B-factories measure a number of parameters of the B decays that relate to CP violation.
 Thus the time dependent CP asymmetry for $B^0$ decaying into a final state $f$ is given by:
 \begin{eqnarray}\label{eq02}
\mathcal{A}_{f}(t) & \equiv &
\frac{\Gamma(\overline{B}^0(t) \rightarrow f) -
\Gamma(B^0(t) \rightarrow f)}
{\Gamma(\overline{B}^0(t) \rightarrow f) +
\Gamma(B^0(t) \rightarrow f)}
\end{eqnarray}
which is parametrized by \begin{equation}
 \mathcal{A}_{f}(t)=-C_{f}\cos (\Delta m_B t) + S_{f}\sin (\Delta m_B t)
\end{equation}
with \begin{equation}
S_{f}\equiv sin2\beta_f\;,\;\;\;\;C_{f}=\frac{1-
|A(\overline{B}^0(t) \rightarrow f)/A(B^0(t) \rightarrow f)|^2}{1+
|A(\overline{B}^0(t) \rightarrow f)/A(B^0(t) \rightarrow f)|^2}, \end{equation}
 and $\Delta M_{B_d}$ is the $B_d$ mass difference. For charged modes one has
 \begin{equation} \mathcal{A}_{CP} \equiv \frac{\Gamma(B^-
  \rightarrow f^-) - \Gamma(B^+ \rightarrow f^+)}{\Gamma(B^-
  \rightarrow f^-) + \Gamma(B^+ \rightarrow f^+)} \end{equation}
 $C_f$ and $\mathcal{A}_{CP}$ give a measure of direct CP violation. In addition the branching ratios
 have been well measured.
 
 We consider first the decay $B^0\rightarrow J/\Psi K_s$. 
 The BABAR and ELLE measurements give \cite{psi}.
 \begin{equation}sin2\beta_{J/\Psi K_s}=0.734\pm0.055,\,C_{J/\Psi K_s} = 0.052 \pm 0.047\end{equation}
 In the Standard Model for any decay $B^0\rightarrow f$, $sin2\beta_f$ should be close to  $sin2\beta$ of
 the CKM matrix and indeed an evaluation of the CKM $\beta$ (without using the B-factory data)
 gives \cite{s2b}
 \begin{equation}sin2\beta=0.715^{+0.055}_{-0.045}\end{equation}
 in good agreement with Eq.(19). Since $B^0\rightarrow J/\Psi K_s$ proceeds through the tree
 diagram in the SM while SUSY effects begin only at the loop level, one would expect only a
 small SUSY correction to $sin2\beta_{J/\Psi K_s}$, again in accord with Eqs. (19). Note also that the
 smallness of $C_{J/\Psi K_s}$ implies that there is very little direct CP violation.
 
 We consider next the $B\rightarrow \phi K$ decays. These decays begin for both the SM and SUSY at the loop level and
 so deviations from the SM result might occur. Using the recent new data from Belle \cite{bb0} and
 the preliminary analysis of new BABAR data \cite{bb} one has:
 \begin{equation}sin2\beta_{\phi K_s}=-0.15\pm0.33,\,C_{\phi K_S} = -0.19 \pm 0.30\end{equation}

We see that $sin2\beta_{\phi K_s}$ differs from the expected Standard Model result of 
$sin2\beta_{J/\phi K_s}$(given in Eq.(19)) by\footnote{The preliminary analysis of
the new BABAR data is about 1$\sigma$ higher than the published older data\cite{bb1}, are so if
the latter were used the discrepancy would be even larger} 2.7 $\sigma$. In addition BABAR has
measured \cite{bsg1} $\mathcal{A}_{CP}(B^-
  \rightarrow \phi K^-)=(3.9+8.7)$\%.

Further suggestion that there may be a breakdown of the Standard Model comes from the BABAR and
Belle measurements of the branching ratios \cite{bsg1}:
\begin{equation}\mbox{Br}(B^0\rightarrow\phi K_s)=(8.0\pm 1.3)\times 10^{-6} \end{equation}
\begin{equation}\mbox{Br}(B^\pm\rightarrow\phi K^\pm)=(10.9\pm 1.0)\times 10^{-6} \end{equation}

\begin{center}
\begin{figure}[t]
  \scalebox{0.85}[0.85]{\includegraphics{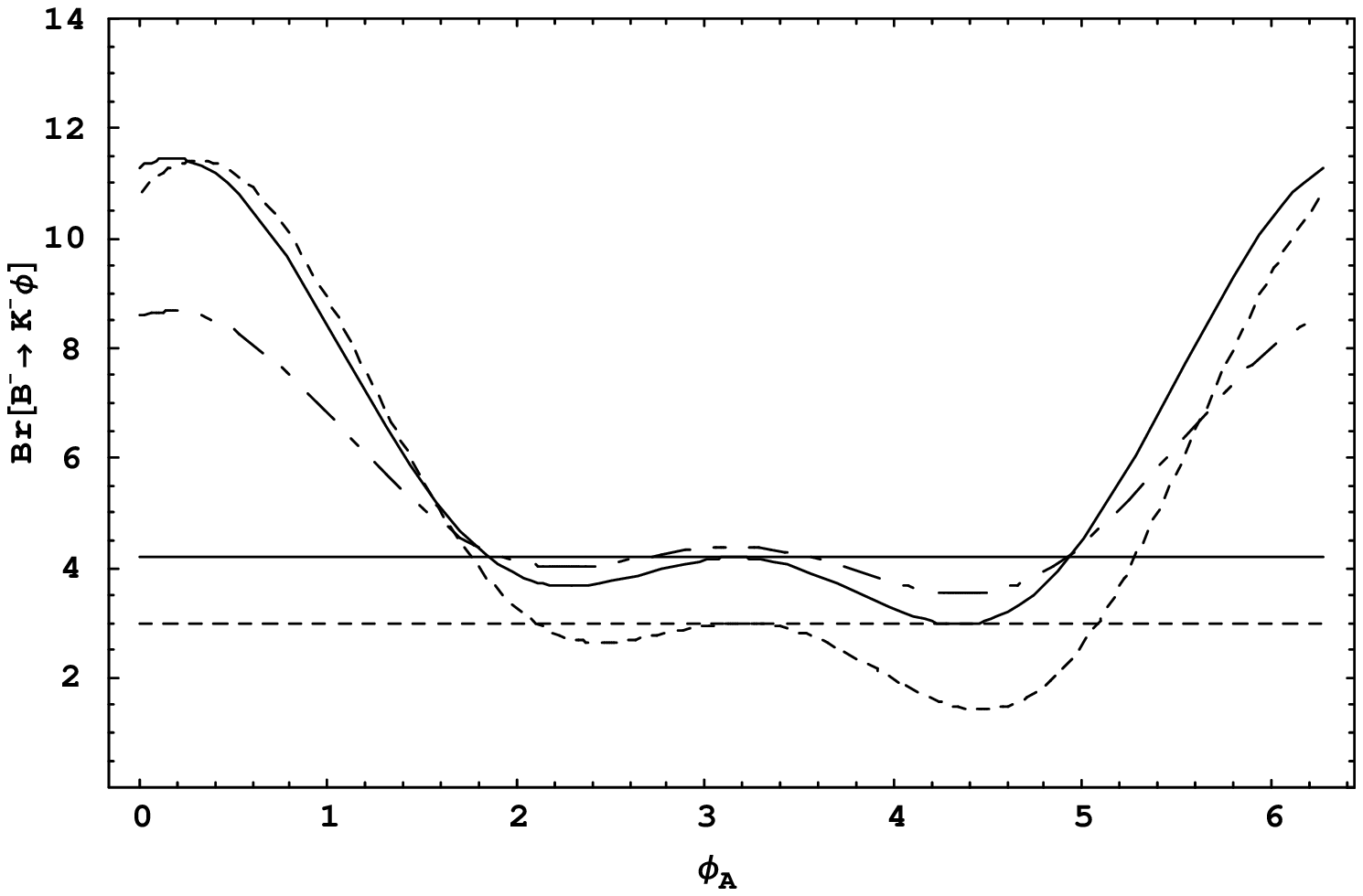}}
  \caption{\label{phi}Branching ratio of $\bpkm$  at $\rho_A = 1$. The solid 
	curve corresponds to $\mu = m_b$, dashed curve for $\mu = 2.5\,\mbox{GeV}$ with 
	$m_s(2\,\mbox{GeV}) = 96\, \mbox{MeV}$ and the dot-dashed curve for 
	$\mu = m_b$ with $m_s(2\, \mbox{GeV}) = 150\,\mbox{MeV}$. The two  
	straight lines correspond to the cases without weak annihilation \cite{we}.}
\end{figure}
\begin{figure}[t]
  \scalebox{0.80}[0.80]{\includegraphics{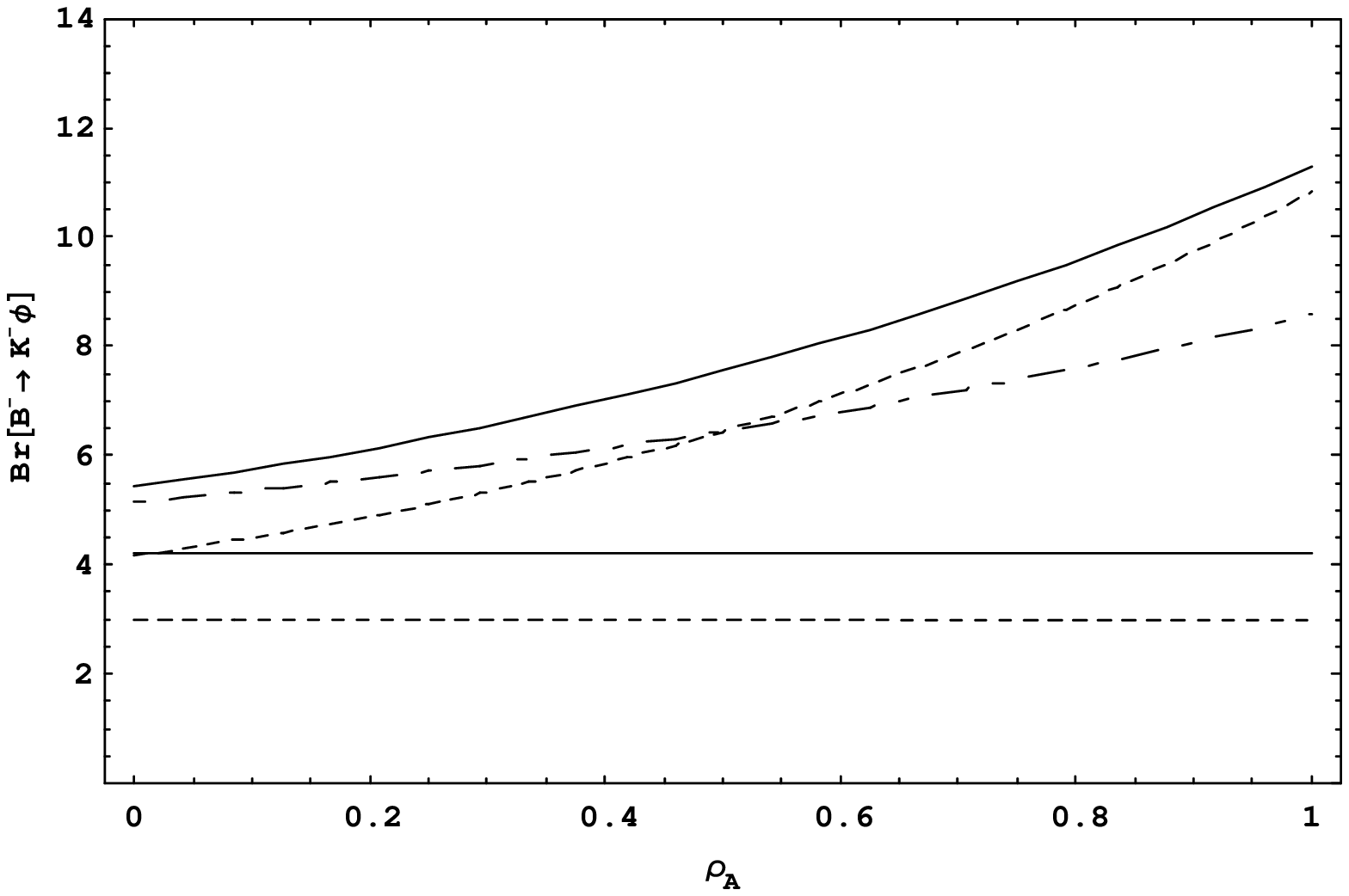}}
  \caption{\label{rho}Branching ratio of $\bpkm$  at $\phi_A = 0$. The solid curve corresponds 
	to $\mu = m_b$, dashed curve for $\mu = 2.5\,\mbox{GeV}$ with $m_s(2\,
	\mbox{GeV}) = 96\, \mbox{MeV}$ and the dot-dashed curve for $\mu = m_b$ with 
	$m_s(2\, \mbox{GeV}) = 150\,\mbox{MeV}$. The two  straight lines correspond to 
	the cases without weak annihilation \cite{we}.}
\end{figure}
\end{center}

While the BBNS analysis is a significant improvement over naive factorization, it is not a
complete theory. The largest theoretical uncertainty comes from the weak annihilation diagram
(where a gluon splits into two final state quarks). These diagrams have divergent end-point
integrals $X_A$ (which are cut off at $\Lambda_{QCD}$) and parametrized by:
\begin{equation} X_A
= (1+\rho_A e^{i\phi_A})\ln
\frac{m_B}{\Lambda_{QCD}},\,\rho_A\leq 1. \end{equation}
Fig. 6 shows the dependence of $\mbox{Br}(B^\pm\rightarrow\phi K^\pm)$ on $\phi_A$ for $\rho_A=1$, and
Fig. 7 the dependence on $\rho_A$ for $\phi_A$=0. We see that the branching ratio over the most
of the parameter space is $\simeq 4\times 10^{-6}$ and does not get large unless $\rho_A$ is
near its maximum ($\rho_A\simeq 1$) and $\phi_A$ is near 0 (or 2$\pi$). However, Figs 6 and 7
also show that when the branching ratio becomes large and is then in accord with the
experimental value of Eq.(23), the weak annihilation diagram dominates the decay amplitude, and
so the theory is least reliable. Thus in the region where the theory is most reliable the
predicted SM branching ratio is too small and in order to obtain a SM value in accord with
experiment, one must go to a region of parameters where the theory is least reliable. (A
similar result holds for $\mbox{Br}(B^0\rightarrow\phi K_s)$.) However, even if one were to do this, it would not resolve the
2.7$\sigma$ discrepancy of $sin2\beta_{\phi K_s}$, since $sin2\beta_{\phi K_s}$ is insensitive to $\rho_A$ and $\phi_A$. Hence that
discrepancy with the SM would still remain.

One can next ask if the SUSY corrections of mSUGRA can resolve the difficulties
of the SM for the $B\rightarrow\phi K_s$ decays. The answer is no. One finds, as
one varies $\tan\beta$, $A_0$, $m_{1/2}$ that $sin2\beta_{\phi K_s}\simeq
0.69-0.74$ which is essentially the same as the SM value and is in disgareement
with the experimental value of Eq.(21)). Also, unless the weak annihilation
diagrams are larger $\mbox{Br}(B\rightarrow\phi K_s)\simeq 4\times 10^{-6}$ for mSUGRA and again is
too small to account for the experimental values of Eqs.(22) and (23). The
reason mSUGRA cannot account for a reduction of $sin2\beta_{\phi K_s}$ from the SM value
is that in mSUGRA the only flavor violating source is in the CKM matrix which
cannot provide enough flavor violation in the $b\rightarrow s$ transition of the $B\rightarrow
\phi K$ decays.

\section{SUGRA With Non-Universal A Terms}
One can ask whether one can add any non-universal soft breaking terms to mSUGRA
to try to account for the apparent disagreement between experiment and the SM in
$sin2\beta_{\phi K_s}$. In a GUT model, at least the SM gauge group must hold at
$M_G$ and so there are only two ways one can enhance mixing between the second
and third generations. one can have non-universal squark masses, $m_{23}^2$, at
the GUT scale or non-universal A terms in the u or d sectors. The first
possibility gives rise to left-left or right-right couplings only, and it was
shown in \cite{bb3} that these produce only a small effect on the $B\rightarrow
\phi K$ decays.  We therefore consider instead $A^{U,D}_{23}$ terms which
produce left-right couplings\cite{we}. We write
\begin{equation}\label{eq14}
A^{U,D} = A_0 Y^{U,D} + \Delta A^{U,D}
\end{equation} Here $Y^{U,D}$ are the Yukawa matrices and so the first term in
the universal contribution. We assume that $ \Delta A^{U,D}_{ij}$ has non-zero
elements only for $i=2$, $j=3$ or $i=3$, $j=2$, and write $\Delta
A^{U,D}_{ij} = |\Delta A^{U,D}_{ij}| e^{\phi^{U,D}_{ij}}$. Tables 1 and 2, for
$\tan\beta=10$ and 40 show that there is a wide range of parameters that can
accommodate the experimental results of Eq. (21) with $\Delta A^{D}_{23}\neq 0$.
We assume here parameters such that the weak annihilation effects are small and
so the theoretical uncertainty is reduced. In spite of this the large
experimental branching ratio of $B\rightarrow \phi K$ are achieved as can be
seen in Table 2. ($\mbox{Br}(B^\pm\rightarrow\phi K^\pm$ is $\sim 10\times 10^{-6}$ for
all entries of Table 1). The parameter space giving satisfactory results for the
case of $\Delta A^{U}_{23}\neq 0$ is more restrictive generally requiring large
$\tan\beta$ and lower $m_{1/2}$ as can be seen in the examples in Table 3.
However, it is still possible for the theory to be within $1\sigma$ of the
experiment with reasonable choices of parameters.
\begin{table}[h]
 \begin{tabular}{|c|c|c|c|c|c|}
 \hline $|A_0|$ & $800$ & $600$ & $400$& $0$ & $|\Delta A^{D}_{23(32)}|$\\
 \hline  $m_{1/2}=300$ & $\ba{c} -0.50 \ea$ & $\ba{c} -0.49 \ea$ & $\ba{c} -0.47 \ea$ & $\ba{c} -0.43 \ea$ & $\ba{c} \sim 50 \ea$\\
 \hline  $m_{1/2}=400$ & $\ba{c} -0.43 \ea$ & $\ba{c} -0.40 \ea$ & $\ba{c} -0.38 \ea$ & $\ba{c} -0.36 \ea$ & $\ba{c} \sim 110 \ea$\\
 \hline  $m_{1/2}=500$ & $\ba{c} -0.46 \ea$ & $\ba{c} -0.46 \ea$ & $\ba{c} -0.44 \ea$ & $\ba{c} -0.31 \ea$ & $\ba{c} \sim 200 \ea$\\
 \hline  $m_{1/2}=600$ & $\ba{c} -0.15 \ea$ & $\ba{c} -0.13 \ea$ & $\ba{c} -0.04 \ea$ & $\ba{c} 0.05 \ea$ & $\ba{c} \sim 280 \ea$\\
 \hline
 \end{tabular}
 \caption{$S_{\phi K_S}$ at $\tan\beta=10$ with non-zero $A^D_{23}$ and $A^D_{32}$ \cite{we}.}\label{DA1}
\end{table}

\begin{table}[h]
 \begin{tabular}{|c|c|c|c|c|c|c|c|c|}
 \hline $|A_0|$ & \multicolumn{2}{|c|}{$800$} & \multicolumn{2}{|c|}{$600$} &
\multicolumn{2}{|c|}{$400$} &
\multicolumn{2}{|c|}{$0$}\\
 \hline  $m_{1/2}=300$ & $\ba{c} -0.40 \ea$ & $\ba{c} 10.0 \ea$ & $\ba{c} -0.38
\ea$ & $\ba{c} 10.0 \ea$ & $\ba{c} -0.33
\ea$ & $\ba{c} 10.1 \ea$ & $\ba{c} -0.05 \ea$ & $\ba{c} 10.0 \ea$\\
 \hline  $m_{1/2}=400$ & $\ba{c} -0.11 \ea$ & $\ba{c} 8.0 \ea$ & $\ba{c} -0.05
\ea$ & $\ba{c} 8.0 \ea$ & $\ba{c} 0.04
\ea$ & $\ba{c} 7.9 \ea$ & $\ba{c} 0.28 \ea$ & $\ba{c} 8.0 \ea$\\
 \hline  $m_{1/2}=500$ & $\ba{c} 0.07 \ea$ & $\ba{c} 6.0 \ea$ & $\ba{c} 0.16
\ea$ & $\ba{c} 6.1 \ea$ & $\ba{c} 0.24 \ea$
& $\ba{c} 6.1 \ea$ & $\ba{c} 0.37 \ea$ & $\ba{c} 6.2 \ea$\\
 \hline  $m_{1/2}=600$ & $\ba{c} 0.37 \ea$ & $\ba{c} 6.2 \ea$ & $\ba{c} 0.44
\ea$ & $\ba{c} 6.2 \ea$ & $\ba{c} 0.49 \ea$
& $\ba{c} 6.2 \ea$ & $\ba{c} 0.58 \ea$ & $\ba{c} 6.2 \ea$\\
 \hline
 \end{tabular}
 \caption{$S_{\phi K_S}$ (left) and $\mbox{Br}[\bpkm] \times 10^6$ (right) at
$\tan\beta=40$ with non-zero $\Delta
A^D_{23}$ and $\Delta A^D_{32}$ \cite{we}.\label{DA3}}
\end{table}
\begin{table}[h]
 \begin{tabular}{|c|c|c|c|c|c|c|c|c|c|}
 \hline $|A_0|$ & \multicolumn{2}{|c|}{$800$} & \multicolumn{2}{|c|}{$600$} &
\multicolumn{2}{|c|}{$400$} &
\multicolumn{2}{|c|}{$0$} & $|\Delta A^{U}_{23(32)}| $ (GeV)\\
 \hline  $m_{1/2}=300$ & $\ba{c} 0.03 \ea$ & $\ba{c} 8.4 \ea$ & $\ba{c} 0.04
\ea$ & $\ba{c} 9.0 \ea$ & $\ba{c} 0.01 \ea$ & $\ba{c} 8.0 \ea$ & $\ba{c} 0.17 \ea$ & $\ba{c} 8.0 \ea$ & $\ba{c} \sim 300 \ea$ \\
 \hline  $m_{1/2}=400$ & $\ba{c} -0.07 \ea$ & $\ba{c} 8.5 \ea$ & $\ba{c} -0.03
\ea$ & $\ba{c} 8.4 \ea$ & $\ba{c} 0
\ea$ & $\ba{c} 7.1 \ea$ & $\ba{c} 0.32 \ea$ & $\ba{c} 6.3 \ea$ & $\ba{c} \sim
600 \ea$ \\
 \hline  $m_{1/2}=500$ & $\ba{c} 0 \ea$ & $\ba{c} 6.5 \ea$ & $\ba{c} 0.07
\ea$ & $\ba{c} 6.4 \ea$ & $\ba{c} 0.18 \ea$ & $\ba{c} 6.0 \ea$ & $\ba{c} 0.44 \ea$ & $\ba{c} 6.1 \ea$ & $\ba{c} \sim 800 \ea$ \\
 \hline  $m_{1/2}=600$ & $\ba{c} 0.27 \ea$ & $\ba{c} 6.1 \ea$ & $\ba{c} 0.30
\ea$ & $\ba{c} 6.1 \ea$ & $\ba{c} 0.35 \ea$ & $\ba{c} 6.1 \ea$ & $\ba{c} 0.51 \ea$ & $\ba{c} 5.9 \ea$ & $\ba{c} \sim 1000 \ea$ \\
 \hline
 \end{tabular}
 \caption{$S_{\phi K_S}$ (left) and $\mbox{Br}[\bpkm] \times 10^6$ (right) at
$\tan\beta=40$ with non-zero $\Delta
A^U_{23}$ and $\Delta A^U_{32}$.\label{UA1}}
\end{table}

In all the above cases ($\Delta A^{U}$ or $\Delta A^{D}$) the direct CP violating effects 
are small i.e. $\mathcal{A}_{\phi K^-}\simeq -(2-3)$\% and
also we find $|\mathcal{A}_{CP}(B\rightarrow X_s\gamma)| \cong (1-5)$\%. We see thus that a non-universal A term can account for the
B-factory results on $B\rightarrow \phi K$ decays.

\section{ Conclusions}
                                                                                
We have surveyed here three developments of the past year that have further
narrowed the parameter space of possible SUSY models: the new WMAP data,
the status of the muon magnetic moment anomaly, and the Belle and BABAR
data on CP violating B decays. We have examined these questions within the
framework of SUGRA GUT models as these have fewer free parameters, and so
are more constrained by the array of other data, the full set of all the
data allowing one to see more clearly what the constraints on SUSY are.

The WMAP data has determined the basic cosmological parameters with great
accuracy, and this new data has reduced the error in the values of the cold
dark matter density by a factor of about four. In the region of parameter
space $m_0$ and $m_{1/2} < 1$ TeV, it implies for mSUGRA (and many SUGRA GUT
models) that the lightest stau is only 5 -10 GeV heavier than the lightest
neutralino (the dark matter candidate). As the data from WMAP (and later
Planck) become more and more accurate, these measurements will effectively
determine $m_0$ in terms of the other mSUGRA parameters, a result that could
be tested later at high energy accelerators.
                                                                                
The CMD-2 reanalysis of their $e^+ - e^-$ data has reduced the disagreement
between the Standard Model prediction of the muon magnetic moment and the
Brookhaven measurements. Thus using this data one now finds a 1.9$\sigma$
disagreement, but using instead the tau decay into 2 $\pi$ final states plus
CVC, there is only a 0.7$\sigma$ discrepancy. However, the matter is still not
clear as the $e^+ - e^-$ data is still in disagreement with the tau data at the
2.9 $\sigma$ level. Future precision data from KLOE and perhaps BaBar using
the radiative return method should help clarify the issue as these
experiments will be able to check the CDM-2 data. The Brookhaven
analysis of their $\mu^-$ data will further reduce the experimental error on the muon magnetic moment.
                                                                                
Perhaps the most interesting new data this year is the Belle and BABAR
measurements of the CP violating parameters and branching ratios of  $B^0
\rightarrow \phi K_s$ and $B^{\pm} \rightarrow \phi K^{\pm}$. Current data for
$S_{\phi K_s}$ is in
disagreement with the Standard Model at the 2.7$\sigma$ level, and further,
SUSY corrections based on mSUGRA give essentially the same results as the
Standard Model. In addition, over most of the parameter space, the Standard
Model and mSUGRA predict branching ratios for these decays that are a
factor of two or more too low.  SUGRA models offer an essentially unique
way of accommodating both anomalies by adding a non-universal contribution 
mixing the second and third generations
in either the up or down quark sectors to the cubic A soft breaking terms
at $M_G$. If added to the down quark
sector, this term would also lower the CP odd Higgs mass opening up a new
region in the $m_0 - m_{1/2}$ plane satisfying the relic density
constraint.  (Dark matter detector signals would not be increased,
however.) If added to the up quark sector it would lower the stop mass
again giving a new region consistent with dark matter, and making the 
$\tilde t_1$
more accessible to colliders. Thus if this new data is further confirmed,
it would indicate the first breakdown of the Standard Model and perhaps
point to the nature of physics at the GUT and Planck scales, as string
models with this type of non-universality can be constructed.

\section{Acknowledegement}
                                                                                
This work was supported in part by a National Science Foundation Grant
PHY-0101015 and in part by the Natural Sciences and Engineering Research
Council of Canada.

%

\end{document}